# A Decentralized Shared CAV System Design and Application

Sayed Mehdi Meshkani[a,∗], Bilal Farooq[a]

[a]*Laboratory of Innovations in Transportation (LiTrans), Ryerson University, Canada*

**Abstract**

In this study, we propose a novel heuristic two-step algorithm for shared ridehailing in which users can share their rides with only one more user. The algorithm, which is centrally formulated, starts with matching users and creating a set of passenger pairs in step 1 and is followed by solving an assignment problem to assign passenger pairs to the vehicles. To solve the problem of high computational time in dynamic ride-matching problems, we propose a distributed system which is based on vehicle to infrastructure (V2I) and infrastructure to infrastructure (I2I) communication. To evaluate the distributed system's performance, we compare it with the proposed centralized ridehailing algorithm. Both centralized and distributed systems are implemented in a micro-traffic simulator to assess their performance and their impact on traffic congestion. Downtown Toronto road network was chosen as the study area. Based on our obtained results, the service rate of the distributed system was 91.59% which is close to 95.80% in the centralized system. However, the distributed system yielded much lower computational time compared to centralized. Furthermore, the scalability of the distributed system was shown by testing it on a small network and comparing with the entire network.

*Keywords:* Shared on-demand mobility, ride-matching, shared ridehailing, distributed ride-matching,

## 1. Introduction

With the growth in world population and rapid development in urbanization, the demand for transportation is continuously growing (Shen et al., 2016; Tafreshian and Masoud, 2020). To meet this rising demand with high service quality and reduce the negative impacts of transportation on society and environment, such as road congestion and emissions, more sustainable forms of transportation need to be devised. Shared mobility services such as shared ridehailing and ridesharing can be a promising approach as they can potentially decrease the number of vehicles needed to satisfy the mobility needs of participants and provide them with a point-to-point high level of service. The reduction in the number of cars on the roadways ameliorates the traffic congestion and emission and also lowers the need for parking space in urban areas.

Over the past few years, technological advancements in information and communication technology (e.g., emergence of smartphones, ubiquity of high-speed internet) have facilitated the development of on-demand shared mobility services such as dynamic shared ridehailing and ridesharing (Agatz et al., 2011; Feng et al., 2017). Ride-matching problem is the core of such services in which a rider/driver places

∗Corresponding Author.
  *Email addresses:* smeshkani@ryerson.ca (Sayed Mehdi Meshkani), bilal.farooq@ryerson.ca (Bilal Farooq)

his/her request mostly through a mobile application to be matched with a vehicle/rider, and the rider can opt to share the ride with other riders. Several simulation studies demonstrated that the deployment of such services can significantly decrease the number of vehicles in cities (Zhang et al., 2015; Alonso-Mora et al., 2017; Friedrich et al., 2018). According to Zhang et al. (2015), an on-demand mobility service in the city of Singapore could satisfy the mobility needs of the entire population with only 40% of the current vehicle fleet. Also, Alonso-Mora et al. (2017) suggested that 20% of the current 4 seat taxi cabs in New York City are enough to serve 98% of the current trip demand. However, there are some factors that can adversely affect the individual's willingness to share (WTS). König and Grippenkoven (2020) using an online survey highlighted the importance of travel distance and detour time for participants' willingness to share. They reported long travel times and high detour factors as the main barriers for on-demand shared mobility systems. Their results also indicated that 90% of respondents would prefer to share a ride of 10 min if the discount was 50% or more compared to a private ride. In another study conducted by Lavieri and Bhat (2019), travel time added to the trip to serve other passengers was identified as a great barrier to the use of shared services compared to the presence of a stranger. Furthermore, a discount scheme was developed by Martinez et al. (2015) for a new shared taxi service for city of Lisbon which considers a 40% discount to travelers if the ride is shared with another passenger and 55% if the ride is shared with two other travelers. Regarding the importance of detour time and travel time in individuals' willingness to share, to achieve sustainable benefits from such on-demand shared services and encourage individuals to utilize them, well-designed systems need to be developed that provide a high level of service.

Besides service quality, lack of scalability and high computational time are other challenging aspects of ride-matching problems. To deal with these issues, different decentralized and decomposition approaches have been suggested and used in the literature to convert the matching problem into smaller size sub-problems that are easier to solve (Agatz et al., 2012; Nourinejad and Roorda, 2016; Najmi et al., 2017; Tafreshian and Masoud, 2020). Despite the deployment of different strategies by the aforementioned studies to decompose the ride-matching problem, the control structure of the proposed ride-matching systems is mostly centralized. In centralized control structure, there is a single control entity that stores the whole data and processes it centrally. Prone to failure and higher security and privacy risks for users are some drawbacks of such systems(Zuurbier, 2010; Hawas et al., 2012; Baza et al., 2019; Team, 2019). In contrast, in decentralized systems, there are multiple control agents performing independently. In distributed systems, these agents are able to share information with each other (Zuurbier, 2010; Ge et al., 2017). Operating locally and sharing information in case of need tremendously reduce the computational complexity and makes such systems highly scalable (Hawas et al., 2012; Ge et al., 2017). Moreover, if an outage or a security issue occurs in one component, other components would not be influenced by, and the whole system continues to operate normally (Baza et al., 2019; Team, 2019; Li et al., 2018).

In this study, we propose a novel heuristic two-step ride-matching algorithm for shared ridehailing. To provide users with high-quality service with reasonable detour time, in the proposed ride-matching algorithm, each user can share the ride with only one more user. The algorithm starts with matching users to create a set of passenger pairs in step 1 and is followed by assigning passenger pairs to vehicles in step 2. We define a new index to score each passenger pair based on the proximity of users' current locations and their destinations. A new heuristic algorithm then is designed to match users with each other. Furthermore, to tackle the problems of lack of scalability and high computational time in the centralized ride-matching systems, we design a novel distributed system that works based on infrastructure to infrastructure (I2I) and vehicle to infrastructure (V2I) connectivity. A network of intelligent intersections is utilized as the infrastructure (I2s). These I2s are considered local dispatchers that run the proposed ride-matching algorithm based on the data they receive locally. They also can send/receive the data to/from other I2s in case of necessity.

The rest of this paper is organized as follows. In Section 2, we briefly review the relevant literature on ridehailing and ridesharing services. Section 3 introduces the ridehailing system settings, the details of two-step ridehailing algorithm and the distributed system. Section 4 presents the description of the case study, results and discussions. Finally, Section 5 concludes our findings and provides some directions for future research.

## 2. Background

This overview is divided into two parts. In the first part, the literature on dynamic ride-matching systems, including ridesharing and ridehailing is reviewed, and in the second part, we review the literature on distributed ride-matching systems.

### 2.1. Dynamic ride-matching systems

Agatz et al. (2011) proposed a dynamic ridesharing system to optimally match drivers with single riders using rolling horizon approach with as late as possible matching policy. They considered maximizing distance saving as the ridesharing objective function. In another study, Agatz et al. (2012) pointed out that centralized ridesharing matching may not be fast enough when applying on real-size instances. They suggested decentralized approaches such as partitioning geographic study region as one of the solutions. Shen et al. (2016) presented a Filter-and-Refine method to reduce the size of the ridesharing problem. They partitioned the road network utilizing a grid and then filtered the requests based on a spatio-temporal index. Pelzer et al. (2015) proposed a partition-based ridesharing algorithm to split the road network to distinct regions in order to reduce the solution space. They used an agent-based approach to make matching, and a match was completed if the rider's destination was part of the driver's route. Najmi et al. (2017) presented a clustering method based on the participants' spatial positions to create smaller problems that were faster to solve and then solved the assignment problem to match riders and drivers. Alonso-Mora et al. (2017) developed a high-capacity ridesharing algorithm through a multi-step procedure. First, they created pairwise request-vehicle shareability graph. Next, a graph of feasible trips and the vehicles that can serve them was computed. Finally, they solved an integer linear problem to assign the vehicles to the passengers. Simonetto et al. (2019) proposed a ridesharing system based on a federated architecture in order to assign vehicles to the passengers. In their proposed system, first vehicles were filtered based on their proximity to pick up location and then a linear assignment problem for a batch of requests was solved to assign vehicles to the passengers. Tafreshian and Masoud (2020) presented a graph partitioning-based method based on a one-to-one ride matching to decompose the original ridesharing problem into sub-problems which are easier to solve. They also developed more complex systems, including a system in which a driver share his ride with multiple riders and a system with flexible roles.

*2.2. Decentralized/Distributed ride-matching systems*

In this section, the literature related to distributed/decentralized ride-matching systems is reviewed.

Nourinejad and Roorda (2016) proposed a decentralized auction-based ridesharing algorithm to match riders and drivers. In their algorithm, participants were partitioned based on the geographic position such that drivers considered passengers whose origins or destinations were in the vicinity of the driver's route. Sánchez et al. (2016) presented a fully decentralized peer-to-peer ridesharing system to overcome the privacy concerns and lack of trust among peers in centralized ridesharing systems. Their system was based on the co-utility concept. They also proposed a reputation management protocol that enabled peers to trust each other even when they did not interact before. Silwal et al. (2019) presented a survey on taxi ridesharing system architectures. They classified ridesharing systems into static and dynamic and then categorized dynamic systems into three distinctive architectures, including centralized, distributed, and hybrid systems. They also addressed each architecture and mentioned their prose and cons. Furthermore, they introduced the lack of scalability as the main problem of centralized systems. In contrast, hardware dependant and the need to supporting network were stated as the issues in distributed systems. Baza et al. (2019) referred to a single point of failure and internal and external attackers as some issues of centralized ridesharing systems and proposed a decentralized service based on the public blockchain, named B-ride. They introduced a time-locked deposit protocol to prevent malicious activities. In their proposed system, a driver had to prove that he arrived at the pick-up location on time. The payment was based on the elapsed distance of the driver and rider. They also introduced a reputation model to rate drivers based on their past behavior. Yu et al. (2020) designed an asynchronous distributed ridesharing algorithm to handle efficiently the high dynamic nature of such systems. Their system was based on a local wireless communication between taxis and passengers. They tested their proposed ridesharing system with 4000 taxis to show the scalability of system. By considering 5 min detour tolerance, they reported high taxi occupancy rate and reduced number of taxis on the road. Kudva et al. (2020) illustrated that centralized ridehailing systems are controversial in terms of flexibility, data integrity, and stability. They enumerated some drawbacks of such systems, including inflexible policies and pricing schemes, security issues of passengers' information and transnational data, costly to maintain and manage, highly vulnerable to service attacks, increased response delay to requests due to high bandwidth utilization, and computational complexity. As a solution, they built a decentralized ridehailing system based on blockchain technology, named PEBERS. They developed smart contracts to build and deploy functionalities such as create ride, auto-deposit transfer, cancel and complete ride methods.

Masoud and Jayakrishnan (2017) employed the FCFS strategy for matching, which may not be an appropriate approach, specially for dense urban areas and congested networks. In our proposed ridehailing algorithm, the entire batch of ride requests are considered and matched simultaneously which improves the quality of the matching. Unlike Simonetto et al. (2019), where each vehicle is matched with only one user at each matching time, our proposed ridehailing algorithm can combine two users. This can lead to utilizing the available vehicles more efficiently and improving the quality of service. Although Najmi et al. (2017) and Tafreshian and Masoud (2020) applied clustering and partitioning methods, the structure of their ride-matching systems is still centralized. In contrast, we propose a system in this study that structurally is distributed and is based on I2I and V2I connectivity.

# 3. Methodology

In this section, first, we introduce the settings of the dynamic ridehailing system and its characteristics. We then propose the heuristic dynamic ridehailing algorithm and the distributed ride-matching system.

## 3.1. Dynamic ridehailing system settings

Let $V$ be the set vehicles, $R$ be the set of riders, $P$ be the set of passengers, and $N = R \cup P$ be the set of all users. The purpose of the ridehailing system is assigning online riders to the vehicles and finding corresponding schedules while some constraints need to be met. A rider in the ridehailing system refers to a person who places a ride request mostly through a mobile application. Furthermore, a passenger $p$ refers to a rider that has been assigned to a vehicle. This passenger can be already on-board or still waiting to be picked up. A rider $r \in R$ places his/her order mostly through a mobile application and provides the ridehailing system with his/her origin, destination, and request time which are denoted by $\delta_r$, $\omega_r$, and $t_r$, respectively. The current location of a rider or passenger ($n \in N$) is denoted by $\delta_n^*$. For a rider current location and his/her origin is the same ($\delta_n = \delta_n^*$) while for a passenger they can be different from each other ($\delta_a \neq \delta_a^*$). Vehicles have only two empty seats and an available vehicle is a vehicle that has at least one empty seat.

Moreover, similar to Agatz et al. (2011), we assume that riders provide the ridehailing system with the earliest departure time and latest arrival time. The earliest departure time is the earliest time that riders can depart from their origin, denoted by $e_r$, and the latest time they can arrive at their destination is the latest arrival time, denoted by $l_r$. Based on the rider's earliest departure and latest arrival time, the ridehailing system can estimate the latest time the rider can depart from his origin. The latest departure time is represented by $q_r$ and is calculated by $q_r = l_r - T(\delta_r, \omega_r)$ where $T(\delta_r, \omega_r)$ is the travel time that rider goes directly from his origin to destination. The difference between latest departure time and earliest departure time is called rider's flexibility ($f_r = q_r - e_r$). Without the loss of generality, we assume that riders' travel request time and their earliest departure time are the same ($t_r = e_r$). This assumption makes the system more dynamic and more similar to the real on-demand transportation companies like Uber.

We utilize rolling horizon strategy, suggested by Agatz et al. (2011), to solve the dynamic ridehailing problem. In this strategy, the ridehailing algorithm is run repeatedly at specific times over fixed time intervals referred to as "matching time" $t_m^k$ ($k = 0, 1, 2, ...$) and matching interval ($\Delta^k = t_u^k - t_u^{k-1}$). New riders place their ride requests during most recent matching interval. The system's operator runs the ridehailing algorithm at each matching time and consider both new rider requests and the requests that have not been finalized or expired. A request is finalized when it is assigned to a vehicle. It is expired when the current time $t$ exceeds the request's latest departure time while the request has not been assigned to a vehicle yet. The rolling horizon keeps moving forward until all riders either get matched or expired.

There is a set of constraints $Z$, including a capacity constraint ($z_0$) and two time constraints ($z_1, z_2$). Constraint $z_0$ ensures that a vehicle is available and has at least one empty seat. Constraint $z_1$ ensures that riders should be picked up before their latest departure time ($q_r$). Finally, constraint $z_2$ expresses that riders should be dropped off before their latest arrival time ($l_r$). A vehicle can potentially serve a rider if these constraints $Z$ are satisfied.

*3.2. Two-step ridehailing algorithm*

The proposed ridehailing algorithm includes two steps. Fig. 1 showcases the framework of the algorithm. In the first step of the proposed ridehailing algorithm, users (riders and passengers) are matched and a set of passenger pairs is created. To do so, we design a heuristic algorithm that weights each passenger pair based on a new index we introduce and then selects the best passenger pairs through an iterative process. The second step solves an assignment problem to assign the passenger pairs to the vehicles and finds the optimal vehicle travel path (pick up/drop off order). In the following, These steps will be discussed in detail.

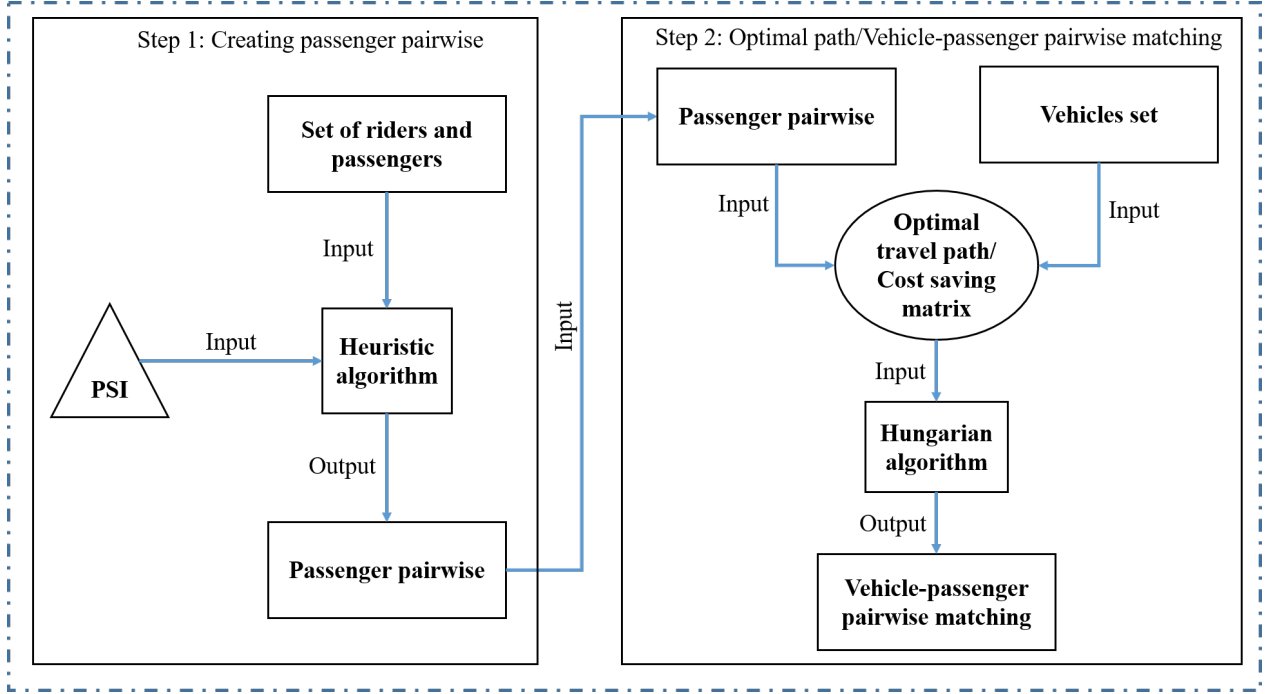

Figure 1: Ridehailing algorithm conceptual model

*3.2.1. Creating passenger pairs*

Let $A$ be the set of all passenger pairs, including riders and passengers: $A = \{(n_j, n_k) : n_j, n_k \in N, (n_j, n_k) = (n_k, n_j)\}$. There are three types of pairs in set $A$: *a)* pairs that both $n_j$ and $n_k$ are riders, denoted by $(r - r)$ *b)* pairs that one of $n_j$ or $n_k$ is rider while the other one is passenger, denoted by $(r - p)$ *c)* pairs that both $n_j$ and $n_k$ are passengers, denoted by $(p - p)$.

As a preprocessing, first we remove the pairs of type *c* in which both $n_j$ and $n_k$ are passengers because they are already assigned to two different vehicles and cannot be matched together. Also, for the pairs of type *a* and *b*, we need to check the shareability. To do so, for the pairs of type *a*, we assume that there is a virtual vehicle located in one of $n_j$ or $n_k$' origin which pick him/her up first and then starts traveling to pick up the other rider. We check whether or not constraint $z_1$, which ensures that riders should be picked up before their latest departure time, is satisfied. For such case, constraint $z_1$ for pairwise $(n_j, n_k)$ is as Eq. 1 and Eq. 2.

$$t + t_{n_j n_k}(\delta_{n_j}, \delta_{n_k}) \leq q_{n_k} \tag{1}$$

$$t + t_{n_k n_j}(\delta_{n_j}, \delta_{n_k}) \leq q_{n_j} \tag{2}$$

where $t$ is current time and $t_{n_j n_k}(\delta_{n_j}, \delta_{n_k})$ is travel time from rider $n_j$'s origin to rider $n_k$'s origin. Pairwise $(n_j, n_k)$ is not shareable if both Eq. 1 and Eq. 2 are violated.

For the pairs of type $b$ in which there is one rider and one passenger, we need to check the constraint $z_1$, only for the rider. Assuming that $n_j$ and $n_k$ are rider and passenger respectively, for the pairwise $(n_j, n_k)$, constraint $z_1$ is as Eq.

$$t + t_{n_k n_j}(\delta^*_{n_k}, \delta_{n_j}) \leq q_{n_j} \tag{3}$$

where $t_{n_k n_j}(\delta^*_{n_k}, \delta_{n_j})$ is travel time from current location of passenger $n_k$ to rider $n_j$' origin. It is needless to say that the preprocessing and checking the pairwise shareability by removing infeasible passenger pairs reduce the size of the problem.

After determining feasible passenger pairs, we introduce a new index, called passenger pair score index (PSI), which scores pairwise $(n_j, n_k)$ based on the proximity of their origins and the proximity of their destinations. PSI is defined as Eq. 4.

$$PSI = \alpha \times \left(1 - \frac{t(\delta^*_j, \delta^*_k)}{1 + t(\delta^*_j, \delta^*_k) + t(\omega_j, \omega_k)}\right) + \beta \times \left(1 - \frac{t(\omega_j, \omega_k)}{1 + t(\delta^*_j, \delta^*_k) + t(\omega_j, \omega_k)}\right) \tag{4}$$

Where $t(\delta^*_j, \delta^*_k)$ is the time distance between users' current locations (rider/passenger) and $t(\omega_j, \omega_k)$ is the time distance between their destinations. $\alpha$ and $\beta$ are parameters which their summation $\alpha + \beta$ is unit. The logic behind PSI is that passengers whose origins and destinations are close to each other can be considered a good match since more likely they would have lower values of waiting time and detour times. From the operator's perspective, this match can reduce the vehicle kilometer traveled (VKT), vehicle travel time, and consequently traffic congestion and emissions.

Fig.2 shows the proximity of two users' origins and their destinations along with their corresponding PSI values. The maximum value for PSI is unit when two users of $n_j$ and $n_k$ have the same current location and also the same destination (Fig.2a) while when their current locations or destinations are different from each other, based on the time distance between their current locations $t(\delta^*_j, \delta^*_k)$ or their destinations $t(\omega_j, \omega_k)$, PSI captures lower values (Fig.2b,c,d).

To match users with each other and create passengers pairs, a greedy heuristic algorithm is presented which has the following steps.

- Step1: Create set $A$, which includes all passenger pairs.
  If the number of users is $|N| = m$, there are $\frac{m \times (m-1)}{2}$ different cases.
- Step 2: Remove infeasible pairs based on the aformentioned preprocessing.
- Step 3: Calculate PSI for feasible pairs, and sort them in a descending order.
- Step 4: Choose pairwise $(p_j, p_k)$ which has the maximum PSI.
- Step 5: Remove all of the pairs from set $A$ which have $p_j$ or $p_k$.
- Step 6: Repeat steps 4 and 5 until the set $A$ is empty.

The output of the proposed heuristic algorithm is the unique passenger pairs.

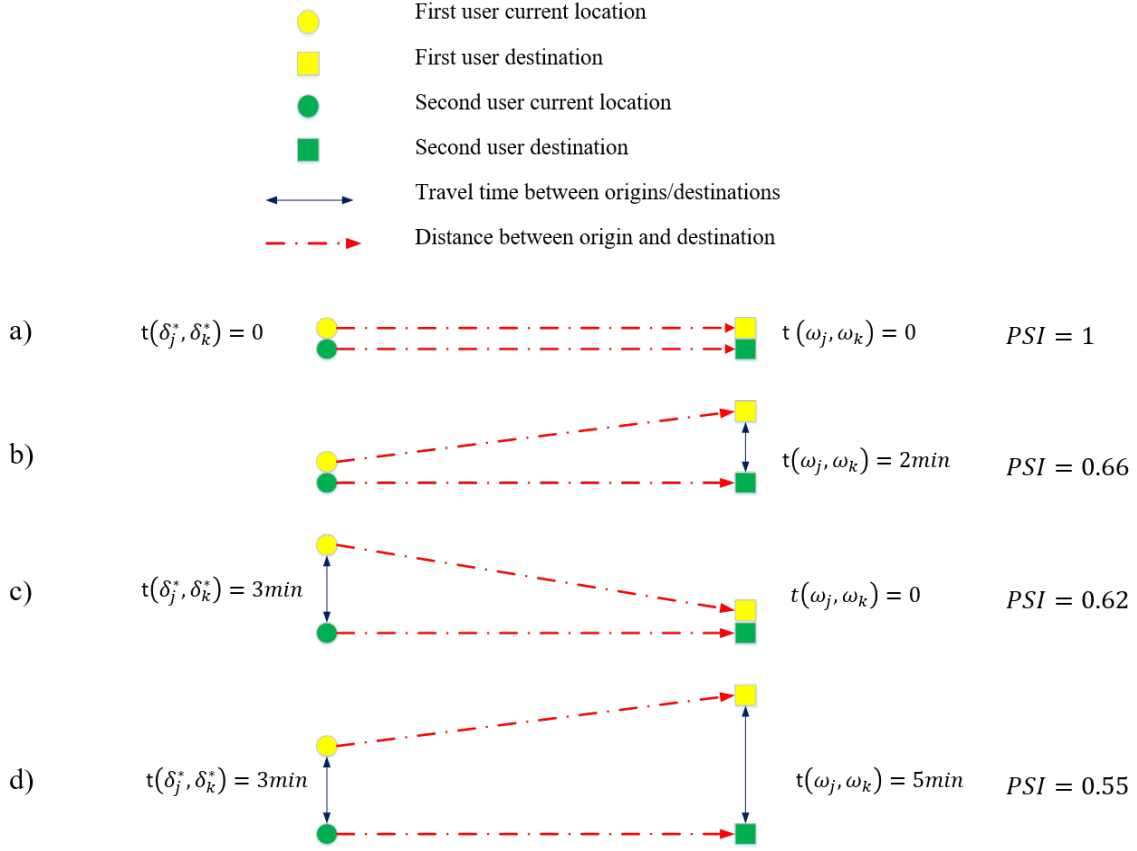

Figure 2: Proximity of users' origins and their destinations

### 3.2.2. Assignment of passenger pairs to vehicles

In the second step of the proposed ridehailing algorithm, first, we obtain the optimal vehicle travel path for the passenger pairs gained from step 1, and then we solve an assignment problem to assign passenger pairs to the vehicles. Maximizing distance saving is one of the objective functions which have been used in the ride-matching problems(Agatz et al., 2011; Nourinejad and Roorda, 2016; Najmi et al., 2017). However, in the congested networks or during peak hours, travel time can better reflect the users' travel cost. Thus, instead of vehicle travel distance saving, we use vehicle travel time saving and maximizing vehicle travel time saving ($VTTS$) is considered as the objective function of the assignment problem.

To acquire the optimal vehicle travel path to pick up and drop off two users of a passenger pair, we enumerate all possible cases. Figure 3 displays six different patterns to pick up and drop off two users. Let *pattset* be the set of patterns and $\Lambda \in$ *pattset* be a pattern. Each pattern has a total travel time, denoted by $TC_\Lambda$, which represents the amount of time a vehicle needs to serve two users. To determine feasible patterns, for each patter, first we check the constraints $z_1$ and $z_2$ to make sure that the vehicle can pick up rider(s) before their latest departure time and drop off users (rider(s)/passenger) before their latest arrival time. For each feasible pattern of a passenger pair ($n_j$, $n_k$), we estimate the vehicle travel time saving ($VTTS$) which is defined as Eq 5.

$$VTTS = T(\delta^*_{n_j}, \omega_{n_j}) + T(\delta^*_{n_k}, \omega_{n_k}) - TC_\Lambda \tag{5}$$

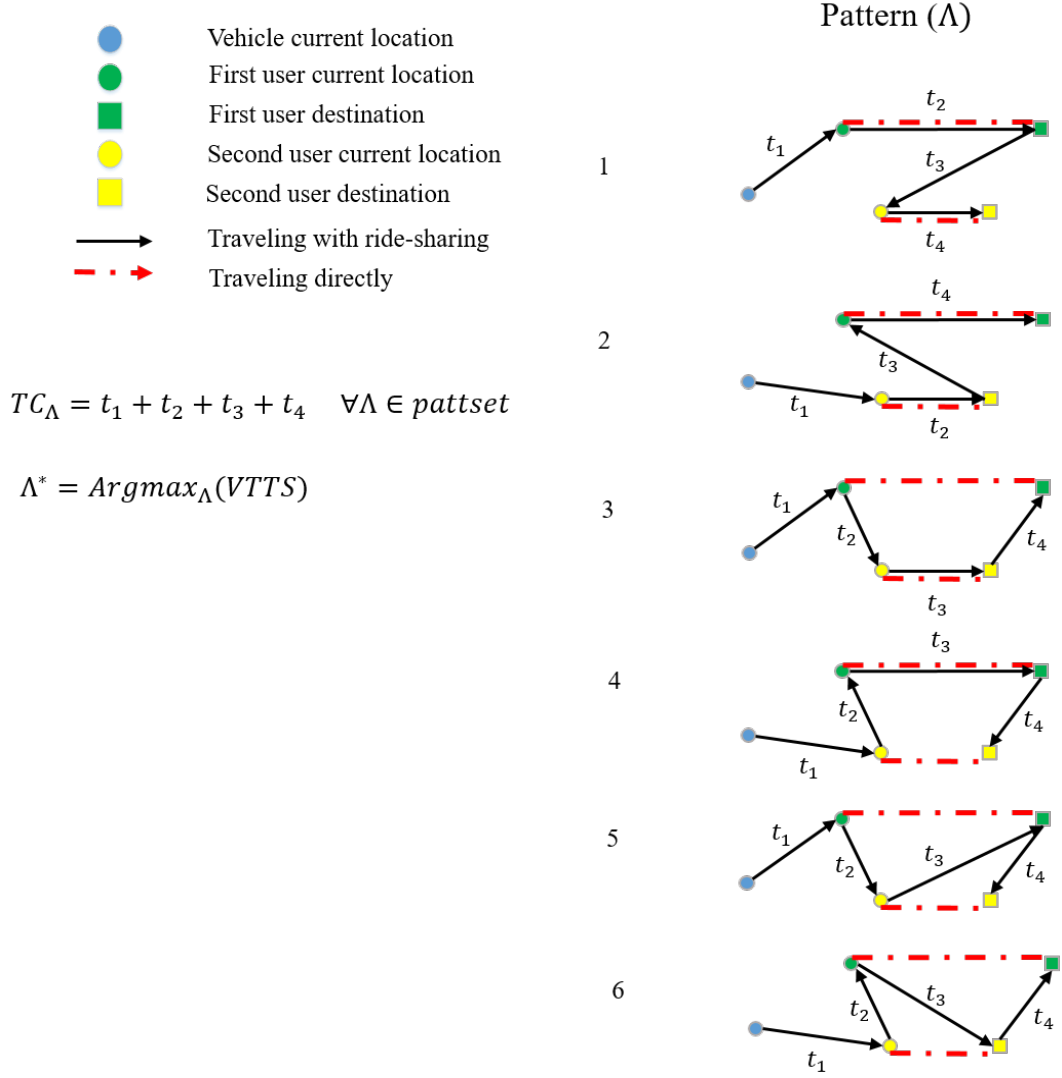

$TC_\Lambda = t_1 + t_2 + t_3 + t_4 \quad \forall \Lambda \in pattset$

$\Lambda^* = Argmax_\Lambda(VTTS)$

Figure 3: Different patterns to pick-up and drop-off two users

where $T(\delta^*_{n_j}, \omega_{n_j})$ and $T(\delta^*_{n_k}, \omega_{n_k})$ are travel time when $n_j$ and $n_k$ goes directly from their current locations to their destinations. Among feasible patterns for each passenger pair, the one with maximum $VTTS$ is the optimal vehicle travel path ($\Lambda^*$) and the associated $VTTS$ is considered its weight, denoted by $\Phi$.

Let $A^i = \{w_1, w_2, ..., w_{m^1}\}$ be the set of modified passenger pairs after removing pairs of type $b$ and $V^i = \{v^i_1, v^i_2, ..., v^i_{n^1}\}$ be the set of idle vehicles. To assign passenger pairs to idle vehicles, an integer linear programming model can be formulated (6). The decision variable $x_{wv^1}$ is 1 if vehicle $v^i$ and pairwise $w$ is matched with each other and 0 otherwise. The objective function (Eq. 6a) aims at maximizing the total travel time saving of the vehicles. Constrains 6b and 6c ensure that each vehicle/pairwise can be matched with only one pairwise/vehicle. Constraint 6d is the binary variable.

$$\max \sum_{w \in A^I} \sum_{v^I \in V^I} \phi_{wv^I} x_{wv^I} \tag{6a}$$

$$\sum_{w \in A^I} x_{wv^I} \leq 1 \qquad \forall v^i \in V^i \tag{6b}$$

$$\sum_{v^I \in V^I} x_{wv^I} \leq 1 \qquad \forall w \in A^i \tag{6c}$$

$$x_{wv^I} = \{0, 1\} \tag{6d}$$

According to Guillaume and Latapy (2006), the proposed assignment problem is equivalent to a bipartite graph matching problem. To solve the matching problem, we use Hungarian algorithm, which is considered one of the effective approaches for solving such problems.

*3.3. Distributed ridehailing system*

In this section, a novel distributed ridehailing system based on infrastructure to infrastructure (I2I) and vehicle to infrastructure (V2I) communication is presented. For the underlying communication and traffic management and without loss of generality, this study uses the framework proposed by Farooq and Djavadian (2019). Their proposed framework, called E2ECAV, is based on a network of intelligent intersections (I2s) and connected and autonomous vehicles (CAVs). I2s are able to process the received local data and share it with each other as well as with CAVs, through I2I and V2I communication, respectively. In our distributed ridehailing system, I2s are considered local dispatchers that are responsible for receiving information from riders and vehicles, running the abovementioned ridehailing algorithm. Also, I2s can send/receive information to/from other I2s through I2I communication. Furthermore, vehicles can share information with I2s through V2I communication. It is worth mentioning that in our distributed system, vehicles only need the capability of connectivity with I2s and being automated is not required. In the proposed distributed system, rider's request is sent to the nearest I2. I2 starts searching process for available vehicles. This search process can be locally on I2's inbound links that we name it search level zero (Fig. 4a). Also, I2 can expand its search space by communicating with its neighbors. Two I2s are neighbors if they are directly connected to each other. For instance, in Fig. 4b, I2 (red spot) has four neighbors. Each neighbor starts searching locally to find available vehicles. Depending on the level at which I2 expands its search space, three more search levels are defined.

In the searching process, vehicles and their associated passengers (for enroute vehicles) may be identified by several I2s. As mentioned, each I2 runs the ridehailing algorithm independent from the other I2s. Thus, a passenger in step 1 of the algorithm may be matched with more than one rider. Similarly, one vehicle in step 2 can be assigned to more than one passenger pair. To avoid this issue, vehicles that are found by several I2s receive the information of PSI and VTTS from related I2s and choose the I2 with the highest PSI or VTTS.

## 4. Case Study and Results

In this section, we briefly introduce the study area and explain how we synthesized the demand. The parameter settings for the simulation of the centralized ridehailing algorithm as well as distributed system on micro-traffic simulator are described. To evaluate the performance of the distributed system, we compare it with the centralized ridehailing algorithm as the base case and discuss on the obtained results. Finally, to assess how changing various variables and ridehailing algorithm parameters affect the performance of the centralized and distributed system, a detailed sensitivity analysis is conduced.

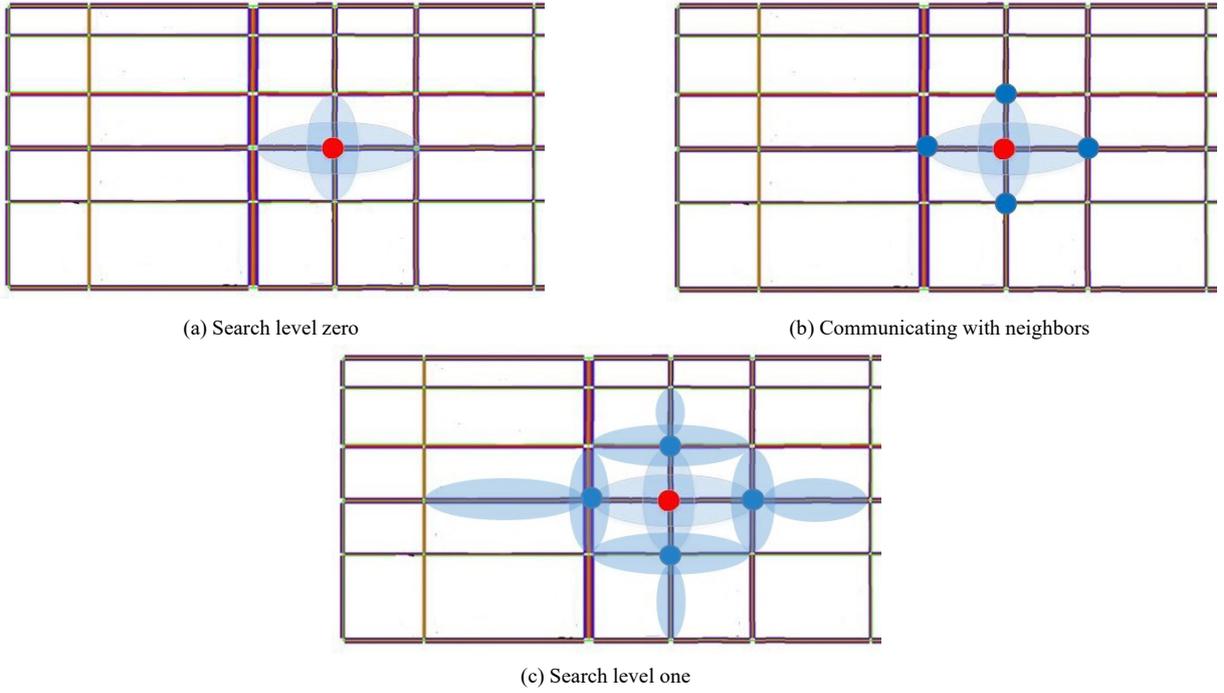

(a) Search level zero    (b) Communicating with neighbors

(c) Search level one

Figure 4: Searching process in the distributed ridehailing system

*4.1. Case Study Implementation*

We used the road network of Downtown Toronto as the study area. One of the reasons for selecting this network is that it faces recurrent congestion during morning and afternoon peak periods. Fig. 5 showcases a small network (marked in red) which consists of 76 nodes/intersections, 223 links with the size of 0.70km x 2.61km, and the entire network (large network), which consists of 268 nodes/intersections and 839 links with the size of 3.14km x 3.31km.

We implemented the ridehailing algorithm, distributed ridehailing system and Downtown Toronto road network in MATLAB and applied them on an in-house agent-based micro-traffic simulator (Djavadian and Farooq, 2018). The dynamic demand loading period in this study was 7:45am-8:00am (15 minutes) in the morning peak period. To simulate the demand, we used 2011 Transportation Tomorrow Survey (TTS) travel data for city of Toronto by applying 2018 growth factor. The simulated demand is time-dependent exogenous Origin-Destination (OD) demand matrices which is based on 5 minutes intervals.The demand within 5 minutes was distributed randomly using a Poisson distribution. The total number of trips in the loading period for the large and small network was 5,487 and 3477 trips, respectively. To select shared vehicles demand, $(20 \pm 5)\%$ of total demand were randomly extracted while the rest of the demand was assumed to travel by their own single-occupancy private vehicles. In this study, we did not have any fleet size optimization, and the size was set exogenously.

Despite the dynamic demand loading period was 15 minutes, the simulation time lasted until all users either arrived at their destination or left the system. It was assumed that link-level space mean speed can be monitored, which was used by the routing agent to provide dynamic travel-time based shortest paths.

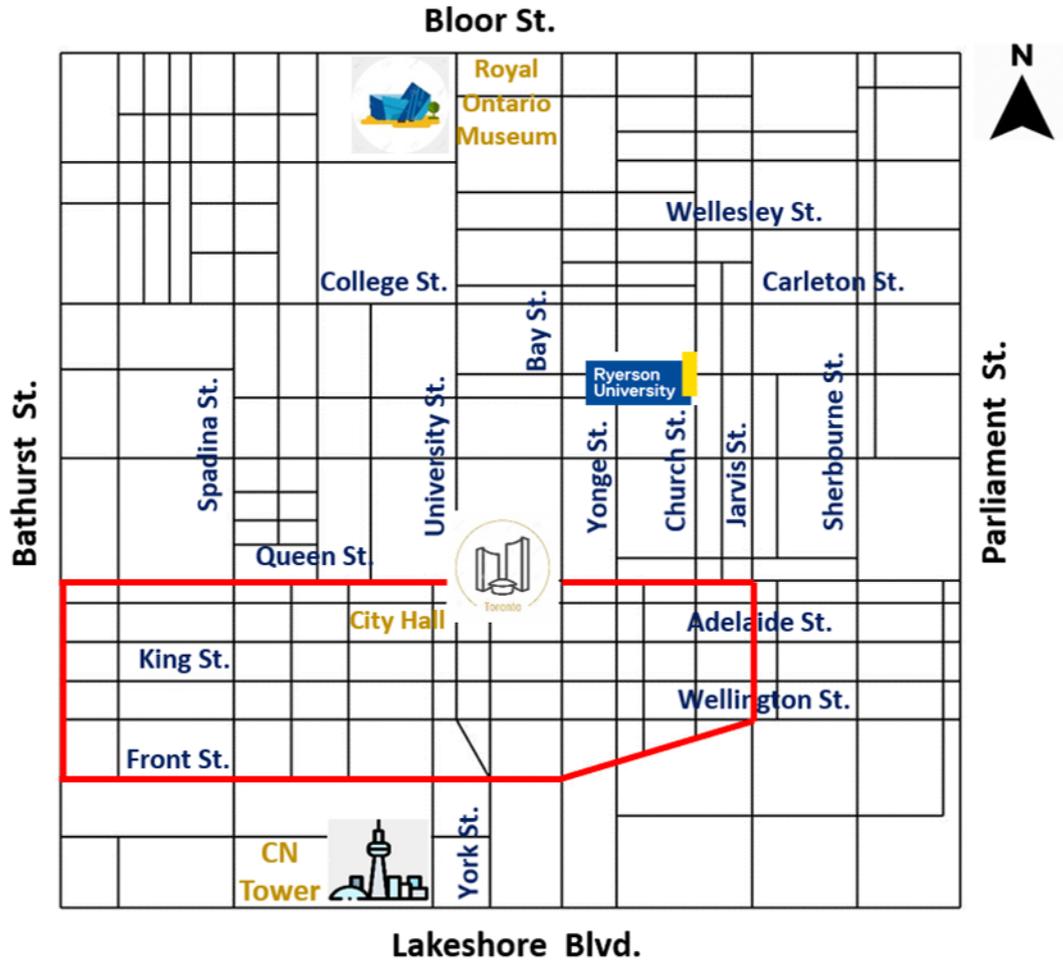

Figure 5: Downtown Toronto street network

The assumption is based on the fact that downtown Toronto already has enough sensors installed that can provide a quasi-real-time state of the network. We assumed that users would leave the system after their latest departure time, if they were not assigned to any vehicles. Also, as mentioned in section 3.1, to make the ridehailing system more dynamic, we assumed that the travel request time equals earliest departure time ($t_r = e_r$).

We considered the proposed centralized ridehailing system as the base case. To evaluate the performance of distributed system, we compared the results with centralized system. To this end, both centralized and distributed system were implemented in MATLAB and were applied on the micro-traffic simulator. In both systems, the shared vehicles were distributed proportional to the ride sharing demand such that more shared vehicles were allocated to the locations with more demand. All simulations in this study were run using three computers, including two computers with Core i7-8700 CPU, 3.20 GHz Intel with a 64-bit version of the Windows 10 operating system with 16.0 GB RAM and one computer with Core i7-6700K CPU, 4.00 GHz Intel with a 64-bit version of the Windows 10 operating system with 16.0 GB RAM.

*4.2. Results*

The results in this section include three parts. First, we compared the performance of distributed ridehailing system with the centralized. In the second part, to show the scalability of the proposed distributed system, we examined its performance on both some part of the roadway network and the entire network. Finally, a detailed sensitivity analysis was conducted on both centralized and distributed systems to show the impact of varying parameters and variables on various indicators. We considered eight indicators to measure for both algorithms. Table 1 shows the indicators with the abbreviations and their descriptions.

Table 1: Indicators, their abbreviations and descriptions

| Indicators | Abbreviation | Description |
| --- | --- | --- |
| Service rate (%) | SR | The percent of served ride requests per total requests |
| Average vehicle km traveled | VKT | Km traveled by each shared vehicle |
| Average detour time (min) | DT | The difference between shared ride travel time and direct travel time for a new request |
| Average wait time (min) | WT | The difference between new rider's pick up time and request time |
| Average traffic travel time (min) | TTT | Average all vehicles' travel time |
| Average traffic speed (km/hr) | TS | Average all vehicles' speed |
| Average number of assignments | No. A | Average number of request assignment per shared vehicle over the simulation period |
| Average computation time per matching time (sec) | - | Average time it takes the algorithm is solved at each matching time |

To compare the performance of distributed ridehailing system with the centralized, we created four scenarios by varying search level (00, 01, 02, 03). Figure 6 showcases the results of this comparison over different indicators. The shared vehicles demand was considered to be 20% of the total demand, which came out to be 1097 trips, and the number of shared vehicles (SVs) was 350. The flexibility was assumed to be five minutes ($f = 5min$), and the update interval was $\Delta = 60sec$. Needless to say that in the proposed ridehailing algorithm, the capacity of vehicles is two ($cap = 2$). It is worth mentioning that the simulation run-time for each scenario was between 30 and 36 hours.

Fig. 6a shows the service rate (SR) for different search levels in the distributed and centralized systems. The centralized as the base case had the most service rate compared to different search levels in the distributed system such that for SV=300, it was 91.22% and for SV=350 it raised to 95.80%. One of the reasons is that the centralized system had the full view of the network and received the information of all existing users and vehicles which enabled it to find more users as well as more available vehicles. By increasing search space over different search levels in the distributed system, the service rate for both SR=300 and SR=350 increased and converged to centralized. The reason is that by enlarging the search space, more available vehicles could be found by intelligent intersections and the probability of successful matching increased. For search level 03 in the distributed system with SV=350, the service rate was 91.59%, which was very close to 95.80% in the centralized with a 4.40% difference. Furthermore, it is crystal clear that for both centralized and distributed systems, by rising fleet size, the service rate as expected enhanced.

Fig. 6b reveals the average computational time per matching time for two algorithms. According to the figure, there is a significant difference in computational time between centralized and various search levels of the distributed system. The computational time for the centralized system with SV=350 was 143.42 sec, while this number for search level 03 of the distributed system was 0.4757 sec. The reason for such a huge difference is that in the centralized system, all of the calculations were computed by a single dispatcher, while in the distributed system, these calculations were divided among the total number of intelligent intersections which is 268 (network nodes).

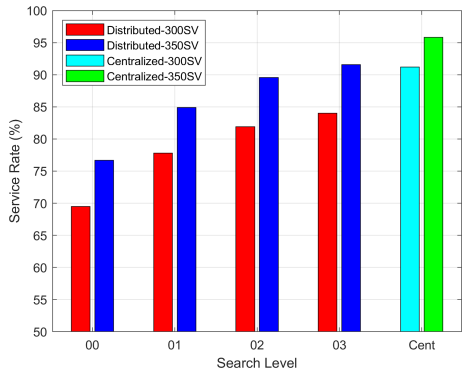
(a) Service rate (%)

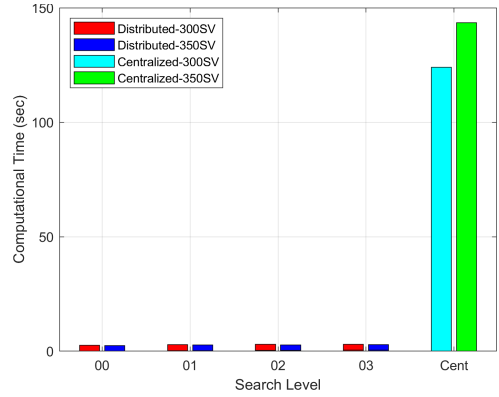
(b) Computational time (sec) per matching time

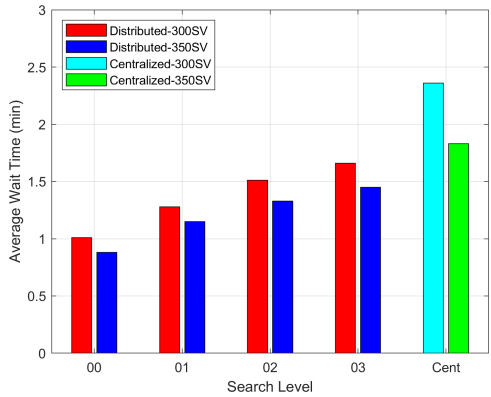
(c) Average wait time (min)

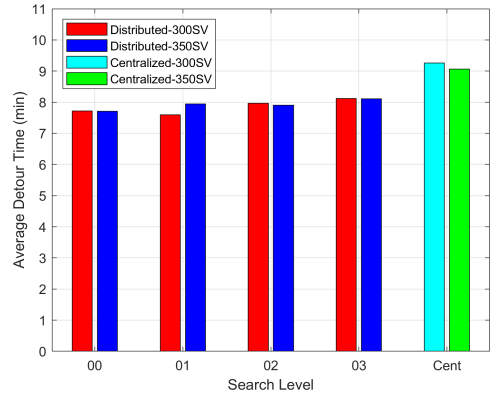
(d) Average detour time (min)

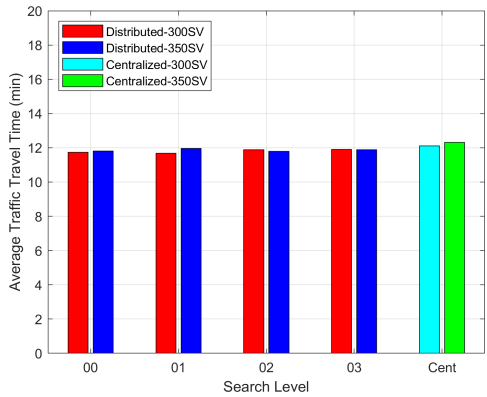
(e) Average traffic travel time (min)

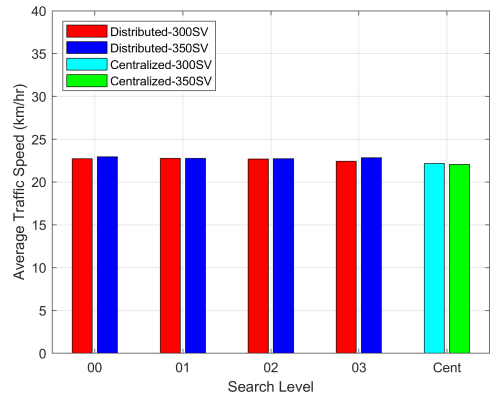
(f) Average traffic speed (km/hr)

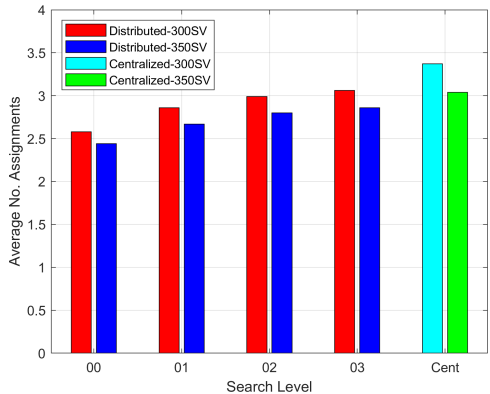
(g) Average No. assignments

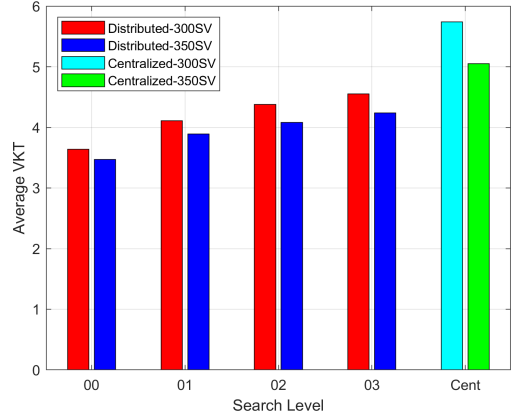
(h) Average vehicle kilometer traveled

Figure 6: Distributed ridehailing system vs centralized: demand=20%, f=5min, cap=2, $\Delta = 60sec$

This result shows the high efficiency of the proposed distributed system in terms of computational time and determines how good they are in reducing the computational burden.

Fig. 6c demonstrates the average wait time (WT) for different scenarios. The centralized system as the base case yielded higher wait time compared to the distributed. This is because the centralized system could find more available vehicles that some of them may be far from the user's origin. Traveling longer distance to pick up the user led to the higher wait time. In the distributed system, by expanding the search space over different search levels, wait time increased. One of the reasons is that in higher search levels, some of the vehicles that were found by an intelligent intersection might be further from it compared to the previous search levels. Thus, the vehicle had to take longer distance to pick up the user. The average wait time for search level 03 with SV=350 showed 20.76% improvement compared to the centralized system which represents the appropriate performance of the distributed system. For both centralized and distributed systems, as expected, raising the number of shared vehicles reduced the wait time.

Fig. 6d showcases the average detour time (DT) over different distributed and centralized scenarios. Detour time is dependent mainly on the proximity of users' origins and their destinations. Based on the obtained results, as expected, the centralized system gave higher detour time than the distributed scenarios. Having a full view of the network increased the probability of matching users with different origins and destinations. In the distributed system, there is a slight difference between various search levels. This is because in both centralized and distributed systems, users' origins and their destinations were matched to the nearest intersections and the demand enters the network from these intersections. Thus, all of the ride requests that were received by each intelligent intersection had the same origin. On the other hand, increasing search space only led to finding more available vehicles, and it does not affect the proximity of users' origins and their destinations. Therefore, by expanding the search space, just a slight increase was observed in the detour time.

Fig. 6e and Fig. 6f represent average traffic travel time (TTT) and average traffic speed (TS) in the network. The traffic travel time and traffic speed for different scenarios in the distributed system were slightly better than centralized such that for search level 03 with SV=350, traffic travel time was 11.80 min, while for the centralized it was 12.30 min which showed 4.00% improvement. Also, for the same scenario, 3.50% enhancement occurred in the traffic speed.

Fig. 6g shows the average number of assignments per vehicle during the simulation period. This indicator for the centralized was higher than the distributed system. This is because the centralized system had access to all information of users and available vehicles. Thus, it was more likely to assign users to available vehicles. In the distributed system, by enlarging the search space, as expected, the number of assignments converged to the centralized. The reason is that the larger search space enabled intelligent intersections to find more available vehicles, leading to the increase in the number of assignments. Also, it is crystal clear from the figure that by increasing the number of vehicles from 300 to 350 with a fixed demand, the average number of assignments per vehicle reduced.

Fig. 6h shows the average vehicle kilometer traveled (VKT) of shared vehicles over the simulation period. The average VKT for the centralized was higher than distributed system. One of the reasons was that in the centralized, vehicles may be called from a longer distance to serve users, which led to a rise in the average VKT. A similar reason can be stated for the distributed. By enlarging the search space, some vehicles located further from the intelligent intersection might be found, thus, they had to travel

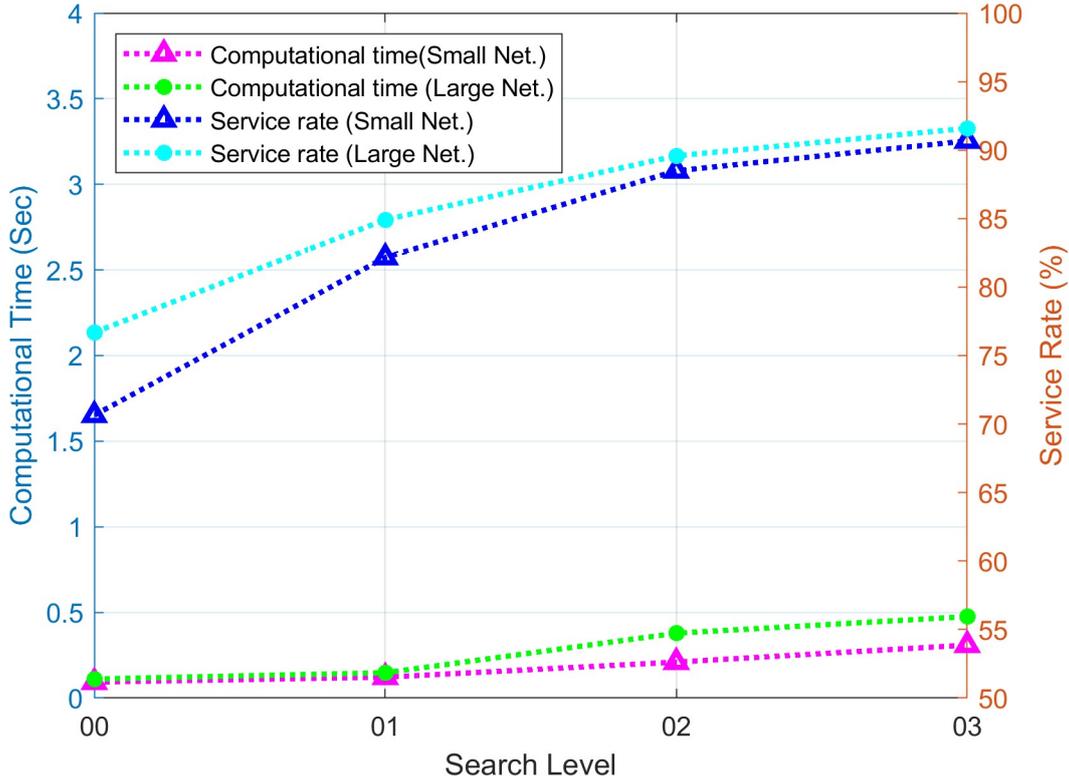

Figure 7: Scalability of the distributed system. Small network: SV demand=695 trips, SV=150, f=5min, cap=2, Δ = 60*sec*. Large network: SV demand=1097 trips, SV=350, f=5min, cap=2, Δ = 60*sec*

longer distance to pick up the users. In search level 03 of the distributed system with 350 shared vehicles, the average VKT was 16% less than the centralized system. Furthermore, by increasing the number of shared vehicles from 300 to 350 in both centralized and distributed systems, each vehicle served fewer users, leading to a reduction in average VKT.

To show the scalability of the proposed distributed ridehailing system, we applied it on the small network in Fig. 5 and compared the average computational time per matching time with the large network. Fig. 7 showcases the performance of the distributed system in terms of the computational time and the service rate for both small and large networks. The shared vehicles demand was 20% of each network's total demand, which was 695 and 1097 trips for the small and large network, respectively. Also, the number of shared vehicles is 150 and 350 for each network. As expected, by enlarging the search space, the service rate increased such that for search level 03, this indicator raised to 90.67% and 91.59% for the small and large network, respectively. However, by scaling up the network, no significant difference was observed in the computational time over various search levels. The average computational time per matching time for search level 03 in the small and large network was 0.3092 sec and 0.4757 sec, respectively. This result revealed that the proposed distributed ridehailing system is highly scalable.

In the last part of the results, a detailed sensitivity analysis was conducted over some variables and parameters of both centralized and distributed systems. Table 2 reports the results obtained by running simulations on both systems. For each system, three scenarios were created by varying demand, fleet size, and flexibility to explore how changing the different parameters affects the system performance. In the first scenario, we considered three fleet sizes of 250, 300, and 350 and 20% as the shared vehicles demand that remained fixed for all the instances. In the second scenario, we considered three demand percentages, including 15%, 20%, and 25% with SV=350, while keeping the other parameters constant.

Table 2: Results for different values of demand, fleet size, and flexibility on centralized and distributed (level03)

| | Fleet Size | Demand (%) | f (min) | SR (%) | VKT | Detour Time (min) | Wait Time (min) | Traffic Travel Time (min) | Traffic Speed (km/hr) | No. Assignments | Computational Time (sec) |
|---|---|---|---|---|---|---|---|---|---|---|---|
| **Centralized** | 250 | 20 | 5 | 80.44 | 5.96 | 9.45 | 2.83 | 11.87 | 22.30 | 3.56 | 194 |
| | 300 | 20 | 5 | 91.22 | 5.74 | 9.26 | 2.36 | 12.11 | 22.18 | 3.37 | 143 |
| | 350 | 20 | 5 | 95.80 | 5.05 | 9.06 | 1.83 | 12.31 | 22.05 | 3.04 | 124 |
| | 350 | 15 | 5 | 96.47 | 3.78 | 8.69 | 1.21 | 12.41 | 21.90 | 2.39 | 87 |
| | 350 | 20 | 5 | 95.80 | 5.05 | 9.06 | 1.83 | 12.31 | 22.05 | 3.04 | 124 |
| | 350 | 25 | 5 | 88.88 | 5.82 | 9.17 | 2.57 | 11.79 | 22.69 | 3.51 | 268 |
| | 350 | 20 | 5 | 95.80 | 5.05 | 9.06 | 1.83 | 12.31 | 22.05 | 3.04 | 124 |
| | 350 | 20 | 10 | 97.99 | 5.23 | 9.10 | 2.11 | 12.32 | 22.02 | 3.11 | 97 |
| **Distributed** | 250 | 20 | 5 | 75.50 | 4.94 | 8.40 | 1.93 | 11.79 | 22.27 | 3.30 | 0.2673 |
| | 300 | 20 | 5 | 84.00 | 4.55 | 8.12 | 1.66 | 11.90 | 22.42 | 3.06 | 0.4535 |
| | 350 | 20 | 5 | 91.59 | 4.24 | 8.11 | 1.45 | 11.88 | 22.84 | 2.86 | 0.4757 |
| | 350 | 15 | 5 | 90.62 | 3.23 | 8.19 | 0.94 | 12.47 | 21.85 | 2.25 | 0.3973 |
| | 350 | 20 | 5 | 91.59 | 4.24 | 8.11 | 1.45 | 11.88 | 22.84 | 2.86 | 0.4757 |
| | 350 | 25 | 5 | 82.52 | 4.86 | 8.22 | 1.74 | 11.58 | 22.85 | 3.23 | 0.5170 |
| | 350 | 20 | 5 | 91.59 | 4.24 | 8.11 | 1.45 | 11.88 | 22.84 | 2.86 | 0.4757 |
| | 350 | 20 | 10 | 96.62 | 4.68 | 8.67 | 1.86 | 12.10 | 22.62 | 3.02 | 0.27 |

Finally, in the third scenario, with the demand of 20% and fleet size of 350, two flexibility levels, including 5min and 10min were tested.

In the first scenario of table 2, increasing the number of shared vehicles, as expected, led to a rise in the service rate such that for SV=350, this indicator reached 95.80 % and 91.59% for the centralized and distributed, respectively. One of the reasons is that with more shared vehicles while the demand was fixed, the centralized/distributed system could find more available vehicles and therefore, more users were assigned to the vehicles. Furthermore, by existing more vehicles in the network, the fixed demand was distributed among more vehicles, leading to a reduction in the average number of assignments per vehicle and the VKT. More available vehicles predictably enhanced the wait time and slightly improved the detour time because each vehicle served fewer users over the simulation period and passengers waited less to be picked up. Traffic travel time in the centralized system slightly increased while in the distributed system, no significant difference was observed. Finally, in the centralized system the average computational time per matching time reduced with the rise in the number of vehicles. This is because at each matching time because of existing more vehicles, more new users were assigned to the vehicles and less users were left for the next matching times. In other words, at each matching time, the most portion of the users were new users that made their requests over the most recent matching interval, and less portion made their requests at the previous matching times.

In the second scenario in table 2, increasing the demand with a fixed number of shared vehicles as expected reduced the service rate, whereas it increased the VKT and the number of assignments. This is because more demand was supposed to be served by the fixed number of shared vehicles. Therefore, each vehicle had to transport more users, which increased the VKT and the number of assignments. Furthermore, as a result of serving more users with a fixed number of shared vehicles, users had to wait more to be picked up, which adversely affected the wait and detour time. As mentioned, the total demand of the network is fixed (5,487 trips). Thus, by increasing the shared vehicles demand, the number of private vehicles travelling in the network reduced, leading to a slight enhancement in the traffic travel time and traffic speed on both systems. Finally, by raising the demand, more users existed at each matching time, and therefore the computational time increased.

In the third scenario in table 2, two flexibility levels of 5 and 10 min were compared with each other. As expected, the service rate for $f = 10$ was higher than $f = 5$ around 2.00% and 5.00% for centralized and distributed systems, respectively. The reason was that with $f = 10$, the users kept staying in the system 5 more minutes. Over this 5 min, some shared vehicles could drop off their users and become available to serve new users. As a result of serving more riders, for both systems the VKT and the number of assignments increased for $f = 10$ min compared to $f = 5$. Also, the wait time reported higher values for $f = 10$. This is because the users had to wait more to be assigned and picked up. More flexibility did not have any significant effect on the traffic travel time and the traffic speed. Finally, As reported, the computational time for $f = 10$ is surprisingly less than $f = 5$ in both systems. The reason is that, in the case of $f = 10$, users kept staying in the ridehailing system longer than $f = 5$. As a result, in the case of $f = 10$, the number of matching time was more than $f = 5$, leading to lower computational time.

## 5. Conclusion and future directions

In this study, we proposed a novel heuristic ride-matching algorithm for shared ridehailing problem in which each user can share his ride with only one other user. The proposed algorithm has two steps. In the first step, utilizing a heuristic algorithm and based on a new index (PSI), users are matched with each other, and the set of passenger pairs is created. PSI scores each passenger pair based on the proximity of users' current location and their destinations. In the second step of the algorithm, an assignment problem is solved to assign passengers pairs to the vehicles. Moreover, to resolve the problems of lack of scalability and high computational time in the centralized ride-matching systems, a novel distributed ridehailing system was designed that performs based on infrastructure to infrastructure (I2I) and vehicle to infrastructure (V2I) connectivity. To evaluate the performance of the distributed system, we compared it with the proposed centralized ridehailing algorithm. Both centralized and distributed systems were implemented on an in-house micro-traffic simulator to investigate their performance in the presence of traffic congestion. Downtown Toronto road network was used as the case study.

The results of comparing distributed with centralized system revealed that the service rate for distributed is 91.59% which is very close to 95.80% in the centralized system. However, the computational time per matching time for the distributed system is much lower (0.4757 sec) than the centralized (143.42 sec). Such improvement demonstrates how well utilizing a distributed system can resolve high computational time in centralized ride-matching problems. VKT in the distributed system showed 16% improvement compared to the centralized. Moreover, wait time in the distributed was 20.76% lower than centralized. Furthermore, our results indicated that the distributed system slightly enhanced the traffic travel time and traffic speed by 4.00% and 3.50%, respectively.

We applied the distributed system on the some part of Downtown Toronto network and compared its performance with the entire network to show the scalability of the proposed distributed system. The results indicated that by enlarging the network, no significant increase was observed in the computational time which can confirm the scalability of the proposed distributed system.

To further examine the performance of the proposed distributed systems, a detailed sensitivity analysis was performed over various parameters, including demand, fleet size, and flexibility. The results showed that increasing demand, number of shared vehicles and flexibility despite changing the service rate did not significantly impact the computational time.

Regarding the capability of the proposed distributed system in reducing computational time, we expect that our matching algorithm and the proposed distributed system in this work will be helpful for the design and operations of on-demand shared mobility services in dense urban areas, where congestion is a recurrent issue. Moreover, the proposed distributed system can be useful when the ride-matching system suddenly gets overload and receives many ride requests over a short period of time. However, to implement such distributed ride-matching system, many electrical devices need to be purchased and be planted at roadway network intersections. Covering the entire city with such a system may cost a lot. Therefore, as mentioned, this system is better to be developed for the dense urban areas.

There are also some directions for the future. To improve the performance of the proposed distributed system and enhance the cooperation among intelligent intersections, more advanced methodologies such as game theory can be utilized. Moreover, regarding the high efficiency of the proposed distributed system in congested networks, a hybrid ride-matching system (e.g., combination of centralized and dis-tributed system) can be developed to cover the entire city.

ACKNOWLEDGEMENTS:

We would like to thank Canada Research Chair program, Ryerson University, and CUTRIC for funding the research. We would also like to thank Dr. Melvin Wong, Dr. Ranwa Al Mallah, and Dr. Shadi Djavadian for their input, time, and insightful discussions.